\documentclass{PoS} 
\usepackage{tabularx}
\usepackage{booktabs}
\usepackage[usenames,dvipsnames]{xcolor}

\newcolumntype{C}{>{\centering\arraybackslash}X}
\newcommand{\te}{$\mathrm{T}_\mathrm{90}$ }
\newcommand{\nh}{$\mathrm{N}_\mathrm{H}$ }

\title{Constraining GRB progenitors environment with Swift XRT}

\ShortTitle{Constraining GRB progenitors environment with Swift XRT}

\author{Dounia Saez$^{a,b}$ \& \speaker{Diego G\"otz}$^{ a}$\\
\llap{$^a$}CEA Saclay - DSM/Irfu/Service d'Astrophysique - F-91191 Gif-sur-Yvette, France	 \\
\llap{$^b$}SUBATECH - Universite de Nantes, Ecole des Mines de Nantes, CNRS/IN2P3 - France \\
E-mail: \email{dounia.saez@subatech.in2p3.fr}, \email{diego.gotz@cea.fr}}

\abstract{The characteristics of the Gamma-Ray Bursts (GRBs) environment may reflect the differences in GRB progenitors: long GRBs are expected to be found in high-density star-forming regions of the GRB host galaxies, while short ones may be associated with an older stellar population that may have had the time to travel far from stellar forming regions in potentially lower density regions. The latter is related to the hypothesis that short GRBs are associated to the merging of compact objects (BH-NS or NS-NS). \\
We used the Swift XRT GRB afterglow archive to compare the intrinsic neutral hydrogen column density values for long and short GRBs within the redshift range 0.1- 1.3, performing a coherent analysis, and excluding from our analysis observations with poor statistics, which reduced our sample to 15 short GRBs. \\
While short GRBs effectively show a median absorption value smaller than long ones the result is not statistically significant. In order to increase our sample we added short GRBs without redshift measure, and we assigned them random redshifts in the same range achieving a marginal increase in the statistical difference between long and short GRBs.}

\FullConference{Swift: 10 Years of Discovery,\\
		2-5 December 2014\\
		La Sapienza University, Rome, Italy }

\begin{document}

\section{Introduction}
The Gamma-Ray Bursts (GRBs) are one of the most violent phenomena in the Universe and the question of their progenitors is yet to be fully answered. From the \textit{CGRO/BATSE} data, a classification of the GRB based on their duration was established \cite{Kouvelioutou}. The distribution of the latter is bimodal with a separation around 2 seconds. If the burst has a duration smaller than 2 seconds, it is classified as a short GRB. On the other hand, should the burst lasts more than 2 seconds, it will belong to the long GRB category. This bimodal distribution is expected in the light of the latest models, as short and long GRBs are supposed to originate from different progenitors.

Current models are in favour of a binary merger (BH-NS or NS-NS) for the explanation of the origin of the GRBs. On the other hand, long GRBs are believe to be the results of the death of massive stars - supernovae. Since the launch of \textit{Swift} \cite{Gehrels}, several short GRBs were localized and many of them have X-ray follow up of the afterglow in the 0.3-10 keV range.

For the two different classes of GRBs, we expect to have also different host galaxy, which can provide indication about the nature of the GRB progenitors. Thus, the study of their environment by the mean of the X-ray afterglows with \textit{Swift/XRT} \cite{Evans} is a strong indicator of their surroundings.

\section{Sample selection}
Multiple criteria have been used to build our sample. First of all, short and long GRBs are separated using the \te criterium and the spectral lag \footnote{However, see the recent work by Bernardini et al. \cite{Bernardini} about spectral lag in long and short GRBs}. All GRBs having a \te smaller than 2 seconds and without spectral lag in the prompt emission were classified as short GRB (see 2.1). Extended soft emission can affect the length of \te, so if the former is slightly superior to 2 seconds, the GRB can be classified as a short GRB and its properties studied to lift the ambiguity. Other GRBs with a \te be superior to 2 seconds and a spectral lag, will be studied as long bursts (see 2.2).

For all GRBs selected this way, other criteria are needed to conduct the study: the chosen time interval of the afterglow should bear no or small spectral evolution (constant hardness ratio over the given interval), sufficient statistics and observation time (more than 200 counts) and robust redshift value, preferentially measured in optical wavelength through spectroscopy.

The galactic contribution to \nh in the direction of each GRB used here comes from  Kalberla et al. \cite{Kalberla}.

\subsection{Short GRBs}
From the criteria above, our final sample has 15 GRBs. 9 of them are also present in the Kopac et al. \cite{Kopac} sample, 1 in the D'Avanzo et al. \cite{Avanzo} one and the 5 remaining were not studied yet. The different values of the hydrogen column density ($\mathrm{N}_\mathrm{H}$) between the two former papers and our results are consistent and are reported, as well as the redshift ($z$) and GRB duration ($\mathrm{T}_\mathrm{90}$) in table \ref{table:SampleShort}. \\
\begin{table}[h!]
\begin{small}
\begin{tabularx}{0.5\textwidth}{l c c C C}     
\toprule
GRB      & T90  & z 	 & $\mathrm{N}_\mathrm{H}$Gal.          & $\mathrm{N}_\mathrm{H}$(z)\\    
         & (s)  &   	 & {\scriptsize $(10^{20}$cm$^{-2})$}   & {\scriptsize ($10^{20}$cm$^{-2})$}\\    
\hline                          
050724  & 3.0  & 0.257 & 14.0 								 & $27.7^{+23.0}_{-17.8}$ \\
051221A & 1.4  & 0.546 & 5.7 								 & $21.7^{+8.6}_{-7.9}$\\
061006  & 0.4  & 0.438 & 14.1 								 & $16.0^{+14.1}_{-11.5}$\\
070714B & 2.0  & 0.923 & 6.4 								 & $26.5^{+12.5}_{-11.4}$\\
070724A & 0.4  & 0.457 & 1.2 								 & $51.3^{+22.9}_{-18.6}$\\
071227  & 1.8  & 0.381 & 1.3 								 & $19.1^{+7.8}_{-7.1}$\\
080905A & 1.0  & 0.122 & 9.0 								 & $11.8^{+20.6}_{-11.8}$\\
090510  & 0.3	 & 0.903 & 1.7 							     & $17.4^{+11.5}_{-9.9}$\\
100117A & 0.3  & 0.915 & 2.7 								 & $35.9^{+21.1}_{-17.3}$\\
100816A & 2.9  & 0.805 & 4.5 								 & $25.2^{+11.1}_{-10.1}$\\
080123  & 115  & 0.495 & 2.3 								 & $13.9^{+4.6}_{-4.6}$\\
101219A & 0.6	 & 0.718 & 4.91 								 & $46.6^{+29.5}_{-23.1}$\\
111117A & 0.5  & 1.3	 & 3.7 									 & $40.5^{+42.6}_{-33.8}$\\
120804A & 0.81 & 1.3	 & 9.3 									 & $143.0^{+81.8}_{-61.7}$\\
130603B & 0.18 & 0.356 & 1.93 								 & $28.7^{+8.37}_{-7.16}$\\
131004A & 1.54 & 0.71  & 10.8 								 & $48.2^{+27.5}_{-22.1}$\\
\hline
\end{tabularx}
\end{small}
\
\begin{small}
\begin{tabularx}{0.5\textwidth}{l c c c c}    
\toprule
 GRB 	& T90  & z 		& $\mathrm{N}_\mathrm{H}$Gal. 			& $\mathrm{N}_\mathrm{H}$(z) \\   
 	 	& (s)  &   		& {\scriptsize $(10^{20}$cm$^{-2})$}     & {\scriptsize ($10^{20}$cm$^{-2})$}\\    
 \hline 
051210	& 1.3  & 0.658  & 1.91 									&$27.7^{+22.7}_{-18.6}$\\ 
 		&  -   & 0.387  & 	- 									&$21.1^{+16.2}_{-14.1}$\\ 
\hline 
060801	& 0.49 & 0.658  & 1.35 									&$10.4^{+14.5}_{-10.4}$\\
 		& 	-  & 1.009  & 	- 									&$15.1^{+21.4}_{-22.6}$\\ 
\hline 
070809	& 1.3  & 0.658 	& 6.40 									&{\footnotesize Hostless}\\
 		& 	-  & 0.565  & 	- 									&{\footnotesize GRB}\\ 
\hline 
120305A	& 0.10 & 0.658 	& 11.3									&$40.4^{+47.8}_{-32.0}$\\
 		& 	-  & 0.565  & 	- 									&$35.6^{+40.8}_{-28.0}$\\ 
\hline 
121226A	& 1.0  & 0.658 	& 6.11 									&$266.4^{+98.3}_{-87.3}$	\\
 		& 	-  & 1.237  & 	- 									&$535.4^{+205.2}_{-168.9}$\\ 
\hline 
130912A	& 0.28 & 0.658 	& 12.3 									&$13.1^{+19.6}_{-13.1}$\\
 		& 	-  & 0.496  & 	- 									&$10.6^{+15.9}_{-10.6}$\\ 
 \hline 
\end{tabularx}
\end{small}
        \caption{Sample of short GRBs. The left table is the GRB sample with measured redshifts. The right table is the GRB sample with fixed or random redshifts.}
        \label{table:SampleShort}
\end{table}

\indent In order to add statistics to our sample, 6 more GRB without redshifts have been chosen for this study : 051210, 060801, 070809, 120305A, 121226A and 130912A. Their redshifts are, for each GRB, either picked at random in or fixed at a median value. The fixed value chosen for the redshift is 0.658, the median value of all measured redshifts. The range of the random redshift goes from 0.133 to 1.3 : this correspond to the minimum and maximum values of our initial sample with robust redshifts. The results are in the table \ref{table:SampleShort} as well.

\subsection{Long GRB}
For this case, we decided to stay below a redshift of 1.3, which is the $z_{max}$ for the short GRB sample. We choose to use the results from Campana et al \cite{Campana} for 23 GRBs, Salvaterra et al.\cite{Salvaterra} for 6 GRBs and we analyzed 10 unpublished GRBs. All results are presented in table \ref{table:SampleLong}.

\begin{table}[h!]
\begin{footnotesize}
\begin{tabularx}{0.4\textwidth}{l|cCc}   
 \hline       
GRB & z & $\mathrm{N}_\mathrm{H}$(z) & Ref.\\
    &   & {\scriptsize ($10^{21}$cm$^{-2})$}\\                       
\hline 
050416A 			& 0.654		&$6.1^{+0.5}_{-0.9}$  	& C10 \\
050525A 			& 0.606		&$1.5^{+0.9}_{-0.7}$  	& C10 \\
050803	 		& 0.422		&$2.0^{+0.2}_{-0.3}$  	& C10 \\
050826	 		& 0.297		&$7.1^{+2.0}_{-2.0}$  	& C10 \\
051016B 			& 0.936		&$7.7^{+1.1}_{-1.0}$  	& C10 \\
060218	 		& 0.033		&$5.2^{+0.5}_{-0.5}$  	& C10 \\
060614	 		& 0.13		&$0.33^{+0.18}_{-0.13}$ & S12 \\
060729	 		& 0.543		&$1.4^{+0.2}_{-0.2}$  	& C10 \\
060904B 			& 0.703		&$4.4^{+0.7}_{-1.2}$  	& C10 \\
060912A 			& 0.94		&$3.2^{+1.5}_{-1.3}$  	& S12 \\
061007			& 1.262 		&$4.5^{+0.3}_{-0.3}$  	& C10 \\	
061021	 		& 0.35		&$0.73^{+0.2}_{-0.1}$ 	& S12 \\
061110A 			& 0.758		&$1.1^{+1.2}_{-0.6}$  	& C10 \\
070208			& 1.165		&$8.6^{+3.0}_{-2.6}$  	& C10 \\
070318	 		& 0.840		&$7.1^{+0.7}_{-1.3}$  	& C10 \\
070419	 		& 0.971		&$3.5^{+1.1}_{-1.0}$  	& C10 \\
071112C 			& 0.823		&$0.6^{+0.5}_{-0.5}$ 	& C10 \\
071122			& 1.14		&$2.2^{+1.2}_{-1.2}$ 	& C10 \\
080319B 			& 0.938		&$1.3^{+0.1}_{-0.1}$ 	& C10 \\
080411			& 1.030  	&$4.6^{+0.4}_{-0.4}$ 	& C10 \\
080413B 			& 1.101		&$2.7^{+0.5}_{-0.5}$  	& C10 \\
080430	 		& 0.767		&$3.9^{+0.3}_{-0.3}$  	& C10 \\
080707			& 1.232		&$3.3^{+1.9}_{-1.9}$ 	& C10 \\
080916	 		& 0.689		&$9.0^{+2.0}_{-3.5}$ 	& C10 \\
081007	 		& 0.530		&$6.2^{+0.5}_{-0.5}$ 	& C10 \\
090424	 		& 0.544		&$5.1^{+0.4}_{-0.3}$ 	& C10 \\
091018	 		& 0.97		&$1.0^{+0.0}_{-0.8}$ 	& S12 \\
091127	 		& 0.49		&$0.76^{+0.35}_{-0.5}$	& S12 \\
100621A			& 0.54		&$18.0^{+1.2}_{-1.1}$ 	& S12 \\	
\hline
\end{tabularx}
\end{footnotesize}
\quad
\begin{tabularx}{0.55\textwidth}{l|ccCC}   
\toprule
 GRB & T90 & z & $\mathrm{N}_\mathrm{H}$ Gal.  & $\mathrm{N}_\mathrm{H}$(z)  \\    
     & (s) &   & {\small $(10^{20}$cm$^{-2})$} & {\small ($10^{20}$cm$^{-2})$} \\    
\hline                          
101219B	& 34.0		& 0.5519			& 3.39 		& $ 10.9^{+9.1}_{-8.4}$\\
110715A & 13			& 0.82			& 24.5  		& $ 56.1^{+16.6}_{-19.2}$\\
120422A & 5.35 		& 0.28			& 3.61 		& $ 22.0^{+4.9}_{-4.5}$\\
120729A & 71.5 		& 0.80			& 14.0 		& $ 35.7^{+12.6}_{-10.9}$\\
120907A	& 16.9		& 0.970			& 5.38 		& $ 25.1^{+15.8}_{-14.1}$\\
121211A	& 182		& 1.023			& 0.958		& $ 50.5^{+18.3}_{-16.2}$\\
131103A	& 17.3		& 0.599			& 1.08 		& $ 63.7^{+26.9}_{-24.4}$\\
140318A & 8.43 		& 1.02			& 2.51 		& $ 45.7^{+38.5}_{-30.4}$\\
140506A & 111.1		& 0.889			& 7.51 		& $ 44.6^{+14.5}_{-13.1}$\\
140512A & 154.8		& 0.725			& 9.70 		& $ 13.4^{+6.9}_{-6.3}$\\
 \hline 
\end{tabularx}
        \caption{Sample of long GRBs. The left table is the GRB sample from Campana et al. (C10) and Salvaterra et al. (S12). The right table is the new GRB sample.}
        \label{table:SampleLong}
\end{table}

Two peculiar GRBs were excluded from this sample : 101225A, the Christmas burst and 111209A. Both of them are too long and are exceptions.

\section{Analysis and results}
All the measurements of \nh from the Swift/XRT spectra were done using \textit{Swift burst analyser} and fitting tool XSPEC version 12.7. : a combination of a powerlaw, a local redshifted absorber and the absorption in the line of sight due to our Galaxy were used to fit the spectra.

\subsection{Sample with measured redshifts}
\begin{figure}[h!]
        \centering
        \includegraphics[width=0.75\textwidth]{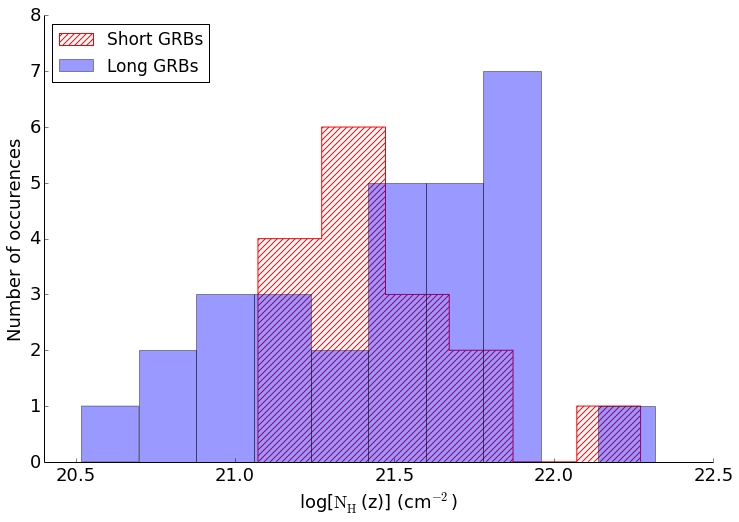}
        \caption{\nh Distribution of GRBs from our sample : red for short ones, blue for long ones}
        \label{figure:HistoSample}
\end{figure}

Our final sample consist of 15 short GRBs at redshift inferior to 1.3 and 39 long ones, at redshift smaller than 1.3. The distributions are plotted in figure \ref{figure:HistoSample}.

The average \nh in this study is 21.46 (in $\mathrm{log_{10}}$) for the short GRBs with a variance of 0.26. Kopac et al. \cite{Kopac} found 21.4 with a variance of 0.1, which is consistent with our sample even though their sample was smaller. In the same way for the long GRBs, we obtained an average value of 21.64 for \nh with a variance of 0.41, whereas Campana et al. \cite{Campana} found 21.9 with a variance of 0.5 but their sample includes GRBs with redshifts larger than 1.3.

A Kolmogorov-Smirnov test was made on our sample to test the probability that the two distributions of short and long GRBs could come from the same initial distribution. We get a probability of 0.5 and a distance of 0.24, so we can not state on a statistical difference between the two distributions.

\subsection{Fixed and random redshift}

As above, the same tests were performed on our second sample, where we included random picked or fixed values of redshift.
\begin{figure}[h!]
        \centering
        \includegraphics[width=0.49\textwidth]{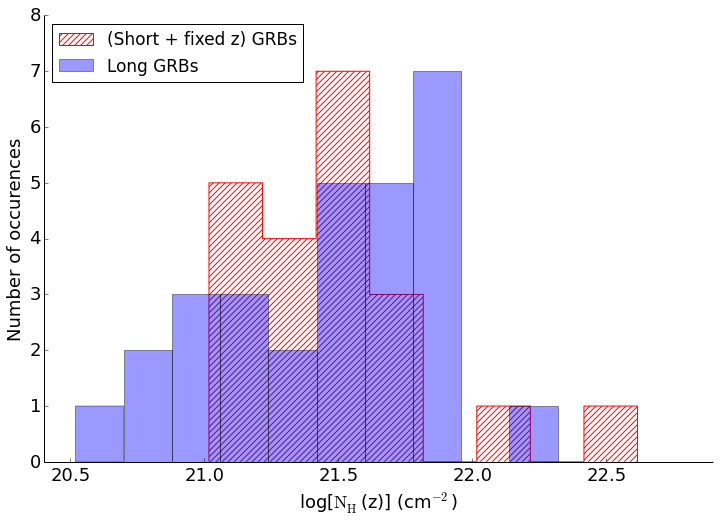}
        \includegraphics[width=0.49\textwidth]{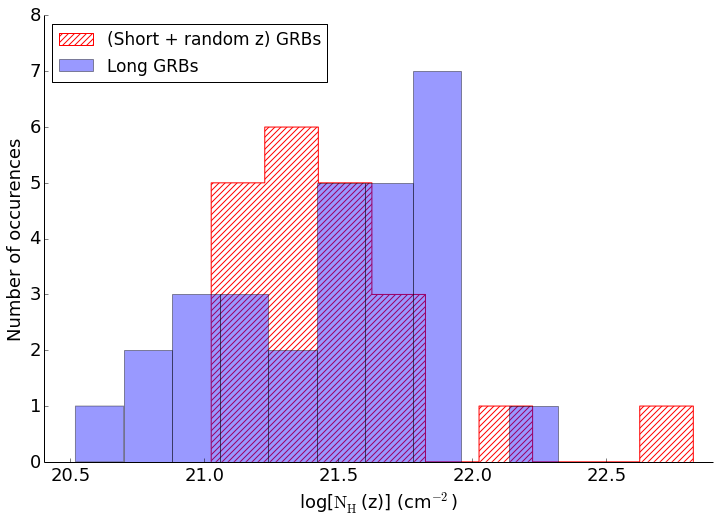}
        \caption{\nh distribution of short and long GRBs with fixed(left) and random(right) redshifts}
		\label{figure:HistoSampleZ}
\end{figure}
For both cases, the Kolmogorov-Smirnov test gave the same probability of 0.51 and the distance of 0.22. The addition of those GRBs to our initial samples did not change the value of the KS probability and thus, we still cannot conclude to any statistical difference between the two distributions.

\section{Discussion and conclusions}

Out study shows that the short GRBs environment is slightly less dense than in the long GRBs ones, as expected by the predictions by the current models. But due to the lack of statistics we cannot fully support these conclusions. More afterglow observations of short GRBs are needed at all redshifts. Better absorption measurements are expected to constrain the theory models. \\
\indent In the future, more Swift/XRT observations and SVOM/MXT, (see G\"otz et al., these proceedings) could help providing the necessary statistics to build a bigger sample of GRBs and thus, determine with better accuracy the difference between the environment of short and long GRBs. \\



\begin{thebibliography}{99}
\bibitem{Kouvelioutou} C. Kouvelioutou et al., \emph{Identification of two classes of gamma-ray bursts}, \emph{ApJ} \textbf{413} (1993) 101
\bibitem{Gehrels} N. Gehrels et al., \emph{Gamma Rays: Bursts, Space Vehicles: Instruments, Telescopes},  \emph{ApJ} \textbf{611} (2004) 1005
\bibitem{Evans} P.~A. Evans et al., \emph{The Swift Burst Analyser I: BAT and XRT spectral and flux evolution of Gamma Ray Bursts}, \emph{A\&A} \textbf{519} (2010) 102
\bibitem{Bernardini} M.~G. Bernardini et al., \emph{Comparing the spectral lag of short and long gamma-ray bursts and its relation with the luminosity}, \emph{MNRAS} \textbf{446} (2015) 1129
\bibitem{Kalberla} P.~M.~W. Kalberla et al., \emph{The Leiden/Argentine/Bonn (LAB) Survey of Galactic HI. Final data release of the combined LDS and IAR surveys with improved stray-radiation corrections}, \emph{A\&A} \textbf{440} (2005) 775
\bibitem{Kopac} D. {Kopa{\v c}} et al., \emph{On the environment of short gamma-ray bursts}, \emph{MNRAS} \textbf{424} (2012) 2392
\bibitem{Avanzo} P. D'Avanzo et al., \emph{A complete sample of bright Swift short Gamma-Ray Bursts}, \emph{MNRAS} \textbf{442} (2014) 2342
\bibitem{Campana} S. Campana et al., \emph{The X-ray absorbing column densities of Swift gamma-ray bursts}, \emph{MNRAS} \textbf{402} (2010) 2492	
\bibitem{Salvaterra} R. Salvaterra et al., \emph{A Complete Sample of Bright Swift Long Gamma-Ray Bursts. I. Sample Presentation, Luminosity Function and Evolution}, \emph{ApJ} \textbf{749} (2012) 68
\end{thebibliography}
\end{document}